\PassOptionsToPackage{dvipsnames}{xcolor}
\documentclass[
  twocolumn,
 showpacs,
 showkeys,
 preprintnumbers,
 amsmath,amssymb,
 aps,
  pra,
  longbibliography,
 floatfix,
 ]{revtex4-2}

\usepackage{mathptmx}

\usepackage{amssymb,amsthm,amsmath}

\usepackage{tikz}
\usepackage {pgfplots}
\pgfplotsset {compat=1.8}
\usepackage{graphicx}

\usepackage{xcolor}

\usepackage{hyperref}
\hypersetup{
    colorlinks,
    linkcolor={blue},
    citecolor={red!75!black},
    urlcolor={blue}
}

\begin{document}

\title{Form of Contextuality Predicting Probabilistic Equivalence between Two Sets of Three Mutually Noncommuting Observables\\[1em]
{\it Dedicated to the memory of Ren\'e Mayet}
}

\author{Mirko Navara}
\email{navara@fel.cvut.cz}
\homepage{https://cmp.felk.cvut.cz/~navara}

\affiliation{Faculty of Electrical Engineering,
Czech Technical University in Prague,
Technick\'a 2,
CZ-166~27 Prague 6,
Czech Republic}

\author{Karl Svozil}
\email{karl.svozil@tuwien.ac.at}
\homepage{http://tph.tuwien.ac.at/~svozil}

\affiliation{Institute for Theoretical Physics,
TU Wien,
Wiedner Hauptstrasse 8-10/136,
1040 Vienna,  Austria}

\date{\today}

\begin{abstract}
We introduce a contextual quantum system comprising mutually complementary observables organized into two or more collections of pseudocontexts with the same probability sums of outcomes. These pseudocontexts constitute non-orthogonal bases within the Hilbert space, featuring a state-independent sum of probabilities. In other words, regardless of the initial state preparation, the total probability remains constant but may be distinct from unity.
The measurement contextuality in this setup arises from the quantum realizations of the hypergraph, which adhere to a specific bound on the linear combination of probabilities. In contrast, classical realizations can surpass this bound. The violation of quantum bounds stems from the inability of classical ontological models, specifically the set-theoretic representation of the hypergraph corresponding to the quantum observables' collections, to adhere to and explain the observed statistics.
\end{abstract}

\keywords{contextuality, two-valued states, quantum states}

\maketitle

\section{Contextuality among mutually noncommuting observables}

Contextuality has various meanings and formalizations in the literature~\cite{Spekkens-04,cabello2021contextuality,svozil-2017-b,Pavicic_2023}.
One of the most common might be in terms of Kochen and Specker's demarcation criterion~\cite[Theorem~0]{kochen1}
concentrating on the separability of any pair of noncommuting observables by two-valued states interpretable as truth assignments.

In this framework, a logico-algebraic structure of propositions is considered~\cite{birkhoff-36},
represented in terms of unit vectors spanning linear subspaces.
These subspaces are constructed through the
orthogonal projections formed by the
summation of dyadic vector products.
The linear span of these subspaces is identified with the logical `or' operator,
the formation of orthogonal subspaces is associated with the `not' operation (or complements),
and set-theoretic intersection corresponds to the `and' operation.

When a collection of propositions is equipped with a
separating set of two-valued states, it can be termed
(quasi)classical. This is due to its set representability, that is, the possibility of embedding  the propositions by a homomorphic (structure preserving) map into a larger Boolean algebra.
On the other hand, if no structure-preserving homomorphic embedding into a larger,
representable Boolean algebra exists, we classify it as contextual as well as (classically) value indefinite.
This, in essence, is Kochen and Specker's demarcation criterion~\cite[Theorem~0]{kochen1}.

However, there are various manifestations of weaker contextuality. Examples include Bell-type inequalities like the Clauser-Horne-Shimony-Holt inequality and the Klyachko inequality, paradoxes such as Hardy's paradox, and configurations like the Greenberger-Horne-Zeilinger setup (which incorporates operator-valued elements and is not exclusively based on elementary propositional operators with eigenvalues of zero and one). What distinguishes these instances are the associated non-Boolean logics.

What unifies them is the potential existence of classical ontologic models, such as set representations, that replicate the respective logics. However, the statistics derived from these ontologies deviate from quantum predictions in terms of probabilities, correlations, and expectations.

To address these situations, Spekkens has classified contextuality in terms of operational equivalence of, say, measurement outcomes, thereby also accommodating statistical forms of contextual behavior~\cite{Spekkens-04}.
Remarkably, these `weaker' statistical forms of contextuality, even though they can be represented using sets and faithfully embedded into Boolean algebras, include complementary observables that are not jointly measurable. When realized in a quantum context, they result in probabilities that differ from those in classical realizations. The significance of these statistical contextual variations lies in their potential for experimental verification
(subject to the assumptions such as counterfactuals),
as opposed to relying on theoretical proofs employing reductio ad absurdum (proof by contradiction).
While the specific concept of contextuality introduced here can be considered within this statistical framework, it retains its independent standing, encompassing novel properties among collections of observables. The same holds true when compared to the varieties of contextuality that can be expressed in and linked to (hyper)graphs discussed previously~\cite{acin-2015,amaral-2018,svozil-2017-b,cabello2021contextuality,Wagner2022Sep,Pavicic_2023}.

In the following sections, we will present a type of contextuality falling within the realm of `weak' contextuality. This classification arises from the fact that its set of observables is both complementary and set-representable. However, classical ontologic models provide statistical predictions that diverge from those derived using quantum probabilities.

These collections of observables represent instances of `extreme' complementarity within the framework of contextuality,
as all propositions associated with them exist in distinct contexts.
As we proceed, we will examine six observables arranged into two triples.
Remarkably, the sum of probabilities for the occurrence of
events within each of these two triples is identical to the sum of probabilities of the other triple.

\section{Pseudocontexts or pseudoblocks}

In what follows, we introduce a new concept: the `pseudoblock' or `pseudocontext' in a hypergraph.
We begin with hypergraphs that are uniformly conformal.
`Uniform' in this context implies that the cardinality of a hyperedge in the hypergraph,
the number of elements of a hyperedge, remains constant for all hyperedges in the hypergraph~\cite[Section~1.1, p.~3]{Bretto-MR3077516}.
`Conformal' means that every maximal clique (although we will primarily focus on uniform configurations, so all cliques are maximal)
is represented by a hyperedge in the hypergraph~\cite[Section~2.4, p.~35]{Bretto-MR3077516}.
Some related graph-theoretical concepts include Greechie~\cite{greechie:71}
and McKay-Megill-Pavi\v{c}i\'{c} (MMP) diagrams~\cite{Mckay2000}.

The hypergraphs under consideration are often motivated by and derived
from configurations of observables in quantum mechanics~\cite{birkhoff-36, kochen1},
or from other algebraic structures such as partition logics~\cite{svozil-2018-b}
and their empirical realizations through initial state identification of finite automata~\cite{e-f-moore} or generalized urn models~\cite{wright}.
At times, hypergraphs are constructed solely to explore `exotic' properties
of specific algebraic structures~\cite{kalmbach-83,nav:91,pulmannova-91}.

As will be elucidated later, to establish a quantum model for any such ad hoc hypergraph of the latter type,
it must possess a faithful orthogonal representation in terms of vector labels.
Similarly, for the development of a classical model that is set-representable,
such as a generalized urn or finite deterministic automaton model,
any ad hoc hypergraph of the latter type must have a partition logic representation.

Any probability distribution on hypergraphs must adhere to the following properties for each hyperedge within the hypergraph:
(i) Exclusivity: This ensures that the probabilities associated with two distinct elements on the same hyperedge are additive.
(ii) Completeness: This requires that the sum of probabilities assigned to all elements within any given hyperedge in the hypergraph equals one~\cite{Gleason, Wright2019}.

For historical reasons, we use the terms `context' or `block' interchangeably to refer to a hyperedge within a hypergraph.
However, we intend to broaden this concept of context or block by considering collections of elements in a hypergraph that:
(i) do not belong to any hyperedge and are, therefore, complementary in quantum-mechanical terms,
(ii) are not necessarily mutually exclusive,
(iii) do not necessarily sum to one in terms of probabilities,
(iv) nevertheless, they have a total probability sum equal to that of other collections of elements in the same hypergraph.
Such collections of elements in a hypergraph will be referred to as `pseudocontexts' or `pseudoblocks'.

We will illustrate this concept with two examples.

\section{Example 1: Generalizations of Firefly Logic}

\begin{figure}
\begin{center}
\resizebox{.26\textwidth}{!}{
\includegraphics{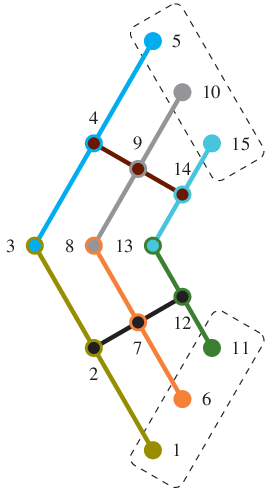}
}
\end{center}
\caption{\label{2023-navara-svozil-Rogalewicz-small}
A hypergraph with two pseudocontexts formed by $\{1,6,11\}$ and $\{5,10,15\}$ marked by dashed boxes.
}
\end{figure}

Consider a 3-uniform (all hyperedges have three atoms or elements) hypergraph
depicted in Figure~\ref{2023-navara-svozil-Rogalewicz-small}~\cite{Navara_1995}.
This example has 15 atoms in 8 contexts.
It was suggested by Ren\'e Mayet as a simplification of the diagram described in the next section. It was used as a cornerstone in~\cite{Mayet_2000}.

\begin{figure}
\begin{center}
\resizebox{.45\textwidth}{!}{
\begin{tabular}{cc}
\includegraphics{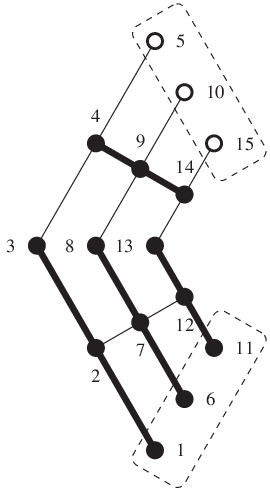}
&
\includegraphics{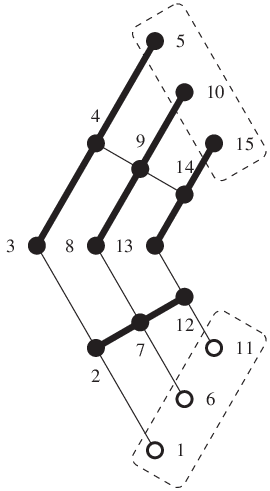}
\\
(a) & (b)
\end{tabular}
}
\end{center}
\caption{\label{2023-navara-svozil-Rogalewicz-proof-small}
Graphical representation of the proof that there are two
triples of elements or atoms $\{1,6,11\}$ and $\{5,10,15\}$ which
are not on a hyperedge and whose respective probability sums are
equal. The `coverings' of the hypergraph depicted in (a) and (b)
include 4 contexts but leave out the elements mentioned.
}
\end{figure}

Assuming exclusivity and completeness there are two triples of elements or atoms
$\{1,6,11\}$ and $\{5,10,15\}$
which are not on a hyperedge and whose probability sums are equal.
The `coverings' of the hypergraph depicted in
Figures~\ref{2023-navara-svozil-Rogalewicz-proof-small}(a) and~\ref{2023-navara-svozil-Rogalewicz-proof-small}(b)
include 4 contexts but leave out the elements mentioned.
Therefore,
\begin{equation}
\begin{split}
\sum_{i=1}^{15} p(i) =
\sum_{i \in
\text{covering (a)}
} p(i) + p(5)+p(10)+p(15) \\
= 4 + p(5)+p(10)+p(15)\\
=\sum_{i \in
\text{covering (b)}
} p(i) + p(1)+ p(6)+p(11) \\
= 4 + p(1)+ p(6)+p(11)
\,
,
\end{split}
\end{equation}
and thus $p(1)+ p(6)+p(11)=p(5)+p(10)+p(15)$, making $\{1,6,11\}$ and $\{5,10,15\}$ pseudocontexts.

\subsection{Representation in terms of sets and vectors}

The hypergraph depicted in Figure~\ref{2023-navara-svozil-Rogalewicz-small} allows both a classical and a quantum representation.

\subsubsection{Quasiclassical representation in terms of partitions of sets}

The hypergraph encompasses a total of 24 two-valued states, which will not be exhaustively enumerated in this article.
There exist two-valued states that are $0$ on all of the observables in the pseudocontexts $\{1,6,11\}$ and $\{5,10,15\}$,
as well as two-valued states that are $1$ on two of them.
As classical probabilities are the convex combinations of all two-valued states~\cite{froissart-81, pitowsky-86}, we obtain bounds for the sum of probabilities $p$ in the pseudocontexts:
\begin{equation}
0 \le p(1) + p(6) + p(11) = p(5) + p(10) + p(15) \le 2.
\label{2023-navara-svozil-bftws-small}
\end{equation}
Therefore, the hypergraph can be used as a false-implies-false gadget for the pseudocontexts $\{1,6,11\}$ and $\{5,10,15\}$.
If the input state is chosen to be triple-0 on one pseudocontext, then the other pseudocontext
exhibits an identical performance. This property is symmetric
with respect to exchange of the pseudocontexts.

A systematic approach for creating set representations of hypergraphs with a separating set of two-valued states, as outlined in Kochen and Specker's demarcation criterion~\cite[Theorem~0]{kochen1}, involves a reverse indexing method that considers all two-valued states~\cite{svozil-2001-eua,svozil-2021-chroma}. Using this method, we can derive a partition logic representation of the pseudocontexts:

\begin{equation}
\begin{split}
{ \bf     b_{1}}  =& \{1,2,3,4,5,6,7,8,9,10\},        \\
{ \bf     b_{6}}  =& \{1,5,6,11,12,15,16,17,19,20\},  \\
{ \bf     b_{11}}  =& \{2,7,9,11,13,15,17,18,21,23\}, \\
{ \bf     b_{5}}  =& \{5,6,7,8,9,10,15,16,17,18\},   \\
{ \bf   b_{10}}  =& \{1,2,3,6,9,11,12,17,20,23\},    \\
{ \bf   b_{15}}  =& \{1,2,4,5,7,11,13,15,19,21\}.
\end{split}
\end{equation}
Classical probability distributions are merely convex combinations of the two-valued states, representing (not necessarily all) `extremal points' within a convex polytope. Consequently, the multiplicities of the entries (and their absence) within the partitions that constitute the two pseudocontexts coincide.

\subsubsection{Quantum representation in terms of vector labels}
\label{2023-navara-svozil-lsFORs}

Lov\'asz introduced a faithful orthogonal representation (FOR) of a graph $G$ with vertices $1, \ldots , n$
by a system of unit vectors
$\{ \vert v_1 \rangle , \ldots , \vert v_n \rangle \}$ in a Euclidean space
``such that if $i$ and $j$ are nonadjacent vertices, then $\vert v_i \rangle$ and $\vert v_j \rangle$ are orthogonal''~\cite{lovasz-79}.
In contradistinction, the more common definition of FORs used here is via the complementary graph $\overline{G}$, such that
if $i$ and $j$ are adjacent vertices---that is, if
they belong to the same edge
also known as
context or block of (maximal) mutually comeasurable observables---then
$\vert v_i \rangle$ and $\vert v_j \rangle$ are orthogonal~\cite{Portillo-2015}.

FORs have a direct quantum interpretation in terms of the orthogonal (that is, self-adjoint) projection operators
$\vert v_i  \rangle \langle v_i \vert $ of the Hilbert space ${\cal H}$
spanning a one-dimensional linear subspace $\vert v_i  \rangle \langle v_i \vert {\cal H}$
which is a formalization of a pure quantum state associated with a unit vector $\vert v_i \rangle$.
The dyadic product of any such vector can be identified as an orthogonal projection operator. This operator, in turn, can be interpreted as a two-valued proposition observable in quantum mechanics.

We  begin by enumerating a FOR obtained through the application of a heuristic algorithm developed by Mc{K}ay, Megill and Pavi{\v{c}}i{\'{c}}~\cite{Pavii2018}:
\begin{equation}
\begin{split}
\vert { \bf   v_{1}} \rangle  & =                        \left( \frac{1}{\sqrt{2}} , \frac{3}{5 \sqrt{2}} , -\frac{2 \sqrt{2}}{5} \right), \\
\vert v_{2}  \rangle  & =                        \left( \frac{1}{\sqrt{3}} , -\frac{7}{5 \sqrt{3}} , \frac{1}{5 \sqrt{3}} \right), \\
\vert v_{3}  \rangle  & =                        \left( \frac{1}{\sqrt{6}} , \frac{1}{\sqrt{6}} , \sqrt{\frac{2}{3}} \right), \\
\vert v_{4}  \rangle  & =   \left( \frac{1}{\sqrt{3}} , \frac{1}{\sqrt{3}} , -\frac{1}{\sqrt{3}} \right), \\
\vert {\bf    v_{5}}  \rangle  & =                        \left( \frac{1}{\sqrt{2}} , -\frac{1}{\sqrt{2}} , 0 \right), \\
\vert {\bf    v_{6}}  \rangle  & =                        \left( \frac{1}{3 \sqrt{2}} , -\frac{8 \sqrt{2}}{15}  , \frac{13}{15 \sqrt{2}} \right), \\
\vert v_{7}  \rangle  & =                        \left( \frac{1}{\sqrt{2}} , \frac{2 \sqrt{2}}{5} , \frac{3}{5 \sqrt{2}} \right), \\
\vert v_{8}  \rangle  & =                        \left( -\frac{2}{3} , \frac{1}{3} , \frac{2}{3} \right), \\
\vert v_{9}  \rangle  & =                        \left( \frac{1}{\sqrt{2}} , 0 , \frac{1}{\sqrt{2}} \right), \\
\vert {\bf    \bf  v_{10}} \rangle  & =   \left( \frac{1}{3 \sqrt{2}} , \frac{2 \sqrt{2}}{3} , -\frac{1}{3 \sqrt{2}} \right), \\
\vert {\bf    v_{11}} \rangle  & =                        \left( -\frac{1}{\sqrt{14}} , \frac{9 \sqrt{\frac{2}{7}}}{5} , \frac{1}{5 \sqrt{14}} \right), \\
\vert v_{12} \rangle  & =                        \left( \frac{1}{\sqrt{6}} , \frac{\sqrt{\frac{2}{3}}}{5} , -\frac{11}{5 \sqrt{6}} \right), \\
\vert v_{13} \rangle  & =                        \left( \frac{4}{\sqrt{21}} , \frac{1}{\sqrt{21}} , \frac{2}{\sqrt{21}} \right), \\
\vert v_{14} \rangle  & =                        \left( -\frac{1}{\sqrt{6}} , \sqrt{\frac{2}{3}} , \frac{1}{\sqrt{6}} \right), \\
\vert {\bf    v_{15}} \rangle  & =                        \left( \frac{1}{\sqrt{14}} , \sqrt{\frac{2}{7}} , -\frac{3}{\sqrt{14}} \right).
\end{split}
\end{equation}

The eigenvalues of
$
\vert v_{1} \rangle \langle v_{1} \vert +
\vert v_{6} \rangle \langle v_{6} \vert +
\vert v_{11} \rangle \langle v_{11} \vert =
\vert v_{5} \rangle \langle v_{5} \vert +
\vert v_{10} \rangle \langle v_{10} \vert +
\vert v_{15} \rangle \langle v_{15} \vert
$ are $2$ with the associated eigenvector $\left( 0,-\frac{2}{\sqrt{5}},\frac{1}{\sqrt{5}} \right)$, as well as   $(7 + \sqrt{21})/14\approx 0.827$, and $(7 - \sqrt{21})/14\approx 0.173$,
respectively.

While there is currently no feasible systematic method for the coordinatization of hypergraphs,
we will additionally present an ad hoc analytical approach for generating a continuum of FORs that differ from those obtained heuristically. This method can be extended to establish a coordinatization for an enlarged hypergraph, encompassing a novel `combo' combination of structures discussed in the next section.
This combination poses a greater challenge, prompting us to employ specific  analytically obtained FORs
to facilitate a seamless continuation, leveraging rotational symmetry as a simplifying tool.

Degenerate cases necessitate individual computations for each coordinatization, typically done manually and potentially with the aid of computer algebra. Some of these cases can be anticipated from the hypergraph itself, without explicit reference to the specific vector representation. However, a thorough analysis is indispensable to identify all potential undesirable relations, such as equalities and orthogonalities, in order to describe all singular configurations accurately.

The analytically obtained coordinatization will, to some extent, be two-dimensional:
The `spiral column' of basis vectors forming the two contexts or blocks
$\{4,9,14\}$ and $\{2,7,12\}$
lies on the hyperplane that is parallel to the $x$-$y$ plane, at $z = 1/\sqrt{3}$.
The third dimension is used to `lift' these two-dimensional vectors so that the proper orthogonality relations are satisfied~\cite{svozil-2016-ggs}.

The contexts $\{4,9,14\}$ and $\{2,7,12\}$
can be coordinatized by assuming that they lie on a cone with an angle
$\text{arccos}\sqrt{ 2/3}$ and an axis that we choose, without loss of generality,
as the `north pole' or $z$-axis $(0, 0, 1)$.

Again, without loss of generality, we may choose vertex $4$
to be represented by the unit vector
$\left( \sqrt{\frac{2}{3}}, 0, \frac{1}{\sqrt{3}} \right)$.
Due to orthogonality and the choice of the cone, we thereby fix the positions of the two vertices $14$ and $26$,
which are in the same context as $2$.
This completes the construction of the orthonormal basis representing  $\{4,9,14\}$.

The context $\{2,7,12\}$ can be obtained by rotating the unit vectors $\{4,9,14\}$
around the $z$-axis $(0, 0, 1)$
by an angle~$\alpha$. As a result, these vectors lie on the same cone.
For the sake of finding an instance, the choice $\alpha = \pi / 3$ complies with the requirements.
Thus, we obtain the orthonormal basis representing $\{2,7,12\}$.

Once the contexts $\{4,9,14\}$ and $\{2,7,12\}$ have been assigned vector labels, the other vertices and contexts
are determined, e.g.\ by cross products.
This concludes the construction of the orthonormal bases representing
$\{3,8,13\}$,
$\{5,10,15\}$, and
$\{1,6,11\}$.
Thus, the vector labels for the hypergraph depicted in Figure~\ref{2023-navara-svozil-Rogalewicz-proof-small} are enumerated.

The above construction can always be performed, but it leads to undesired results in several singular cases discussed below.

Excluding symmetrical solutions, we can, without loss of generality, focus our attention on $\alpha$ within the range of $[0,\pi]$. However, we should exclude the case where $\alpha=0$ because, in that scenario, $\{4,9,14\}$ and $\{2,7,12\}$, as well as $\{5,10,15\}$ and $\{1,6,11\}$, would represent the same triples. In this case, the labels for the lower half of the hypergraph depicted in Figure~\ref{2023-navara-svozil-Rogalewicz-small} would be identical to those of the upper half.

In this degenerate case, the construction yields nine vectors, which,  through proper rotation, can be associated with the edges and diagonals of faces of a cube, for instance,
$\big\{( 1/\sqrt2 )(1,-1,0)$, $ (0,0,1)$, $  ( 1/\sqrt2 )(1,1,0)$, $  ( 1/\sqrt2 )(1,0,-1)$, $  (0,1,0)$, $  ( 1/\sqrt2 )(1,0,1)$, $  ( 1/\sqrt2 )(0,1,1)$, $  (1,0,0)$, $ ( 1/\sqrt2 )(0,1,-1)\big\}$. The `cube representation' is always applicable to configurations such as the upper part of the hypergraph, but it is not the sole representation.

Somewhat surprisingly, the other extreme case, $\alpha=\pi$, does not degenerate and results in the desired configuration.

For $\alpha= 2 \pi /3$, the triples $\{4,9,14\}$ and $\{12,2,7\}$ (in this order) are the same and the degenerate construction produces nothing more than these three vectors.

\begin{figure}
\begin{center}
\resizebox{.35\textwidth}{!}{
\includegraphics{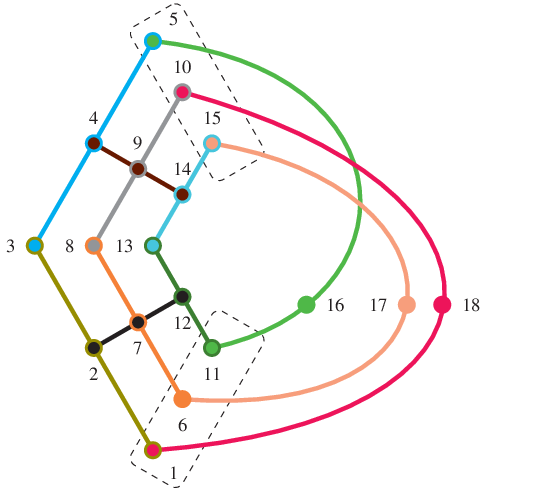}
}
\end{center}
\caption{\label{degenerated}
A degenerated hypergraph for $\alpha_0 \approx 0.886257$ has more orthogonalities than desired.
}
\end{figure}

For all remaining values,
$\alpha\in(0,\pi]\setminus\{ 2\pi/3\}$,
we obtain $15$ distinct vectors satisfying the desired orthogonality relations.
Nonetheless, it is important to acknowledge the possibility of additional orthogonalities that could render our diagram incorrect.
Fortunately, there are only a limited number of vector pairs that require verification for orthogonality.
All linear subspaces form an orthomodular lattice.
It is known that such diagrams do not contain cycles of length $4$, at least in our case of 3-element contexts (for more details, refer to~\cite{kalmbach-83}). Therefore, any pair of vectors representing this `undesirable orthogonality' must have a minimum distance of $4$ in our diagram.
A~typical example is the pair $5$ and $11$, and, up to isomorphism, it appears to be the only one. Using computer algebra
we have determined that this situation occurs for a single value
\begin{equation}
\begin{split}
\alpha_0 &=
2 \text{arctan} \biggl\{ \frac15 \left[- 29
 + 2^\frac{2}{3}  2 \left(75 (69)^\frac{1}{2} +623\right)^\frac{1}{3} \right. \biggl. \\
 & \biggr. \left. \quad
 +\frac{1}{3} \left(538272-64800 (69)^\frac{1}{2}\right)^\frac{1}{3}
 \right]^\frac{1}{2} \biggr\}
\approx 0.886257.
\end{split}
\end{equation}
For potential future reference, we present the corresponding diagram for $\alpha_0$ in Figure~\ref{degenerated}.
In this degenerate case, the hyperdiagram obeys exclusivity (but not completeness) in its respective pseudocontext because it serves as a true-implies-false gadget for the two remaining elements in its pseudocontext.

For all values of $\alpha$ other than those mentioned earlier, we acquire vector coordinatizations
and thus FORs of the diagram depicted in Figure~\ref{2023-navara-svozil-Rogalewicz-small}.
These coordinatizations are nonisomorphic.

\section{Example 2: Generalized false-implies-false and true-implies-true gadget hypergraphs}

\begin{figure}
\begin{center}
\resizebox{.45\textwidth}{!}{
\includegraphics{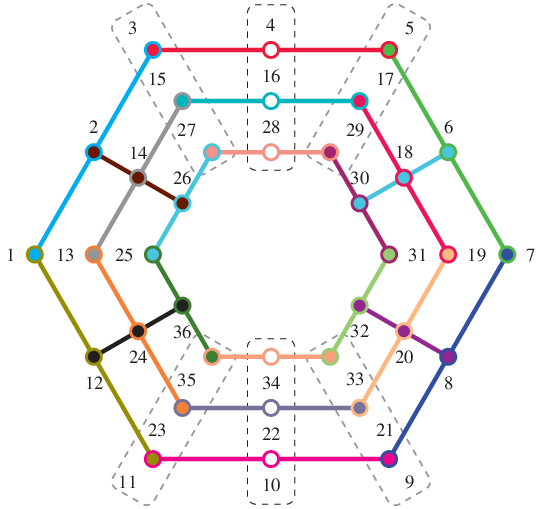}
}
\end{center}
\caption{\label{2023-navara-svozil-Rogalewicz}
Hypergraph of a configuration of observables and contexts containing 36 atoms in 22 contexts.
Pseudocontexts are marked by dashed boxes.
}
\end{figure}

The 3-uniform hypergraph
depicted in Figure~\ref{2023-navara-svozil-Rogalewicz}
is the pasting of two gadget graphs introduced earlier and was first proposed in a letter by Vladim\'ir Rogalewicz~\cite{Rogalewicz90}.
It was the addressee, Ren\'e Mayet, who discovered the potential of this observation and possibilities of its generalization which allowed future results~\cite{Navara_1995}
and~\cite{Mayet_2000}.
This example has 36 atoms in 22 contexts.

\begin{figure}
\begin{center}
\begin{tabular}{cc}
\resizebox{.35\textwidth}{!}{
\includegraphics{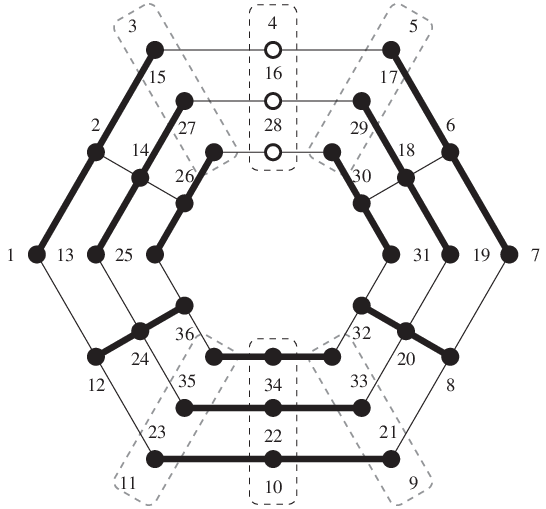}
}
\\ \\ (a) \\ \\
\resizebox{.35\textwidth}{!}{
\includegraphics{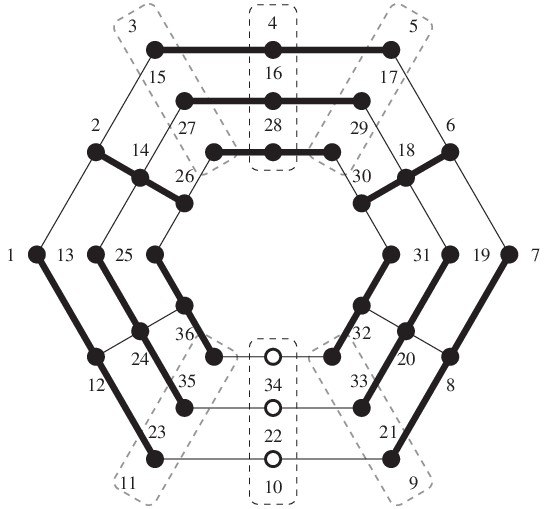}
}
\\
\\
(b)
\end{tabular}
\end{center}
\caption{\label{2023-navara-svozil-proof}
Graphical representation of the proof that there are two triples of elements or atoms
$\{4,16,28\}$
and
$\{10,22,34\}$
which are not on a hyperedge and whose respective probability sums are equal.
The `coverings' of the hypergraph depicted in (a) and (b) include 11 contexts but omit the elements mentioned.
}
\end{figure}

Assuming exclusivity and completeness there are two triples of elements or atoms
$\{4,16,28\}$
and
$\{10,22,34\}$
which are not on a hyperedge and whose probability sums are equal.
The `coverings' of the hypergraph depicted in
Figures~\ref{2023-navara-svozil-proof}(a) and~\ref{2023-navara-svozil-proof}(b)
include 11 contexts but leave out the elements mentioned.
Therefore,
\begin{equation}
\begin{split}
\sum_{i=1}^{36} p(i) =
\sum_{i \in \text{covering (a)}} p(i) + p(4)+ p(16)+p(28) \\
= 11 + p(4)+ p(16)+p(28)\\
=\sum_{i \in \text{covering (b)}} p(i) + p(10)+p(22)+p(34) \\
= 11 + p(10)+p(22)+p(34),
\end{split}
\end{equation}
and $p(4)+p(16)+p(28)=p(10)+p(22)+p(34)$.
One immediate consequence of this result is that two contexts $\{4,16,28\}$ and $\{10,22,34\}$ can be added, but not just one of them.
More precisely, for the coordinatization of the hypergraph in terms of vector labels discussed below,
when the $z$-coordinates of $4$, $16$, and $28$ are $1/\sqrt3$, so are the $z$-coordinates of $10$, $22$, and $34$, and these triples form orthonormal bases.
(In contrast to this, the hypergraph
from Figure~\ref{2023-navara-svozil-Rogalewicz-small}
does not allow the addition of any other context.)

\subsection{Representation in terms of sets and vectors}

The hypergraph depicted in Figure~\ref{2023-navara-svozil-Rogalewicz} allows both a classical and a quantum representation.

\subsubsection{Quasiclassical representation in terms of partitions of sets}

\begin{figure}
\begin{center}
\begin{tabular}{c}
\resizebox{.35\textwidth}{!}{
\includegraphics{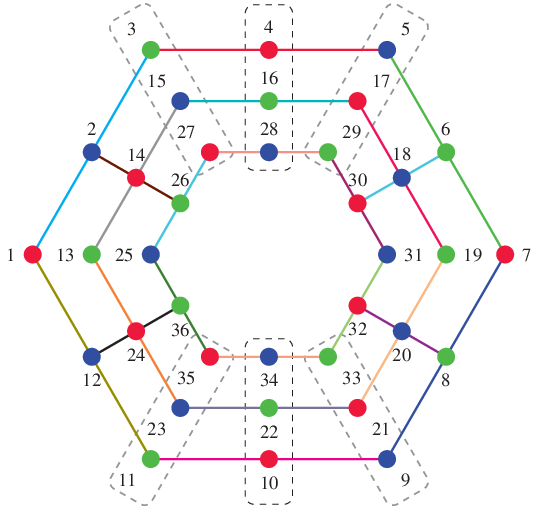}
}
\\ \\ (a)\\ \\
\resizebox{.35\textwidth}{!}{
\includegraphics{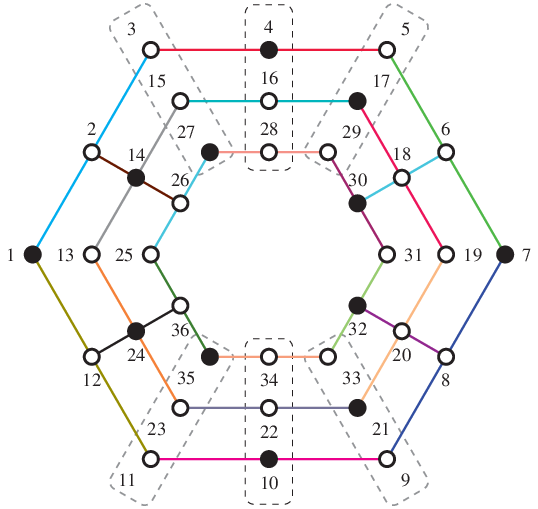}
}
\\ \\ (b)
\end{tabular}
\end{center}
\caption{\label{2023-navara-svozil-chroma}
(a) One canonical~\cite{svozil-2021-chroma} coloring of the hypergraph.
(b) One two-valued state derived from the canonical coloring depicted in (a),
obtained from this coloring by identifying one color `red' with the value `1' and the two remaining colors `green' and `blue' with the value `0'.
}
\end{figure}

The hypergraph encompasses a total of 225 two-valued states, which will not be exhaustively enumerated in this article.
However, it is pertinent to emphasize that the set of states is indeed separable.
In Figure~\ref{2023-navara-svozil-chroma}(a), we present a depiction of
a canonical coloring of the hypergraph~\ref{2023-navara-svozil-Rogalewicz},
where all three colors are represented within each of the blocks, also referred to as cliques.

It is worth noting that in this particular coloring, pseudoblocks or pseudocontexts such as
$\{4,16,28\}$ and $\{10,22,34\}$ also allow a three-coloring.

Identifying two colors with the value $0$ and one color with the value $1$ is a straightforward procedure, as illustrated in Figure~\ref{2023-navara-svozil-chroma}(b). This enables the derivation of one of the 255 two-valued states that is supported
by the hypergraph depicted in Figure~\ref{2023-navara-svozil-Rogalewicz}.

In a classical context, achieving a probability of one for specific observables within the triples ${4,16,28}$ and ${10,22,34}$ is feasible, while registering a probability of zero for the remaining observables. However, this classical scenario contrasts with the quantum mechanical perspective, as the triples are not located
on a shared edge and thus prevent the realization of such probabilities.

Moreover, there exist two-valued states that are $0$ on all of the observables in the pseudocontexts $\{4, 16, 28\}$ and $\{10, 22, 34\}$, as well as two-valued states that are $1$ on all of them. As classical probabilities are the convex combinations of all two-valued states~\cite{froissart-81, pitowsky-86}, we obtain bounds for the sum of probabilities $p$ in the pseudocontexts:
\begin{equation}
0 \le p(4) + p(16) + p(28) = p(10) + p(22) + p(34) \le 3.
\label{2023-navara-svozil-bftws}
\end{equation}

Because of the lower and upper bounds $0$ and $3$, respectively,
the hypergraph can be used as a gadget hypergraph exhibiting a generalized true-implies-false and true-implies-true
sets of propositions in noncontextual hidden-variable theories~\cite{2018-minimalYIYS}:
If, say, the input state is chosen to be triple-$0$ or triple-$1$ on one pseudocontext, then the other pseudocontext exhibits
an identical performance. This property is symmetric with respect to the exchange of the pseudocontexts.
This represents a generalization of the Specker bug~\cite{kochen2,kochen1,specker-ges} and the true-implies-false gadgets of the Hardy type~\cite{Cabello-1996ega,Cabello-2013-Hardylike,svozil-2020-hardy} (for a historical overview, refer to~\cite{2018-minimalYIYS}).

\subsubsection{Quantum representation in terms of vector labels}

Throughout this section, when we refer to `elements' (of edges of the hypergraph) or `points', we are referring to `unit vectors extending from the origin to those points'.

In regard to labeling vertices with vectors, we have not been able to find suitable labels using the previously employed heuristic method.
Therefore, we have extended the analytic strategy for coordinatization used earlier in Section~\ref{2023-navara-svozil-lsFORs}---to find FORs, that is, vertex vector labels obeying mutual orthogonality for the other vertex labels on the same edge---for the hypergraph~\ref{2023-navara-svozil-Rogalewicz} as follows:
The hypergraph will be partitioned into two sides---a `left' and a `right' side---and thereby `cut' along the vertical `axis'
formed by the six elements $\{4, 16, 28\}$ and $\{10, 22, 34\}$.

\begin{figure}
\begin{center}
\resizebox{.46\textwidth}{!}{
\includegraphics{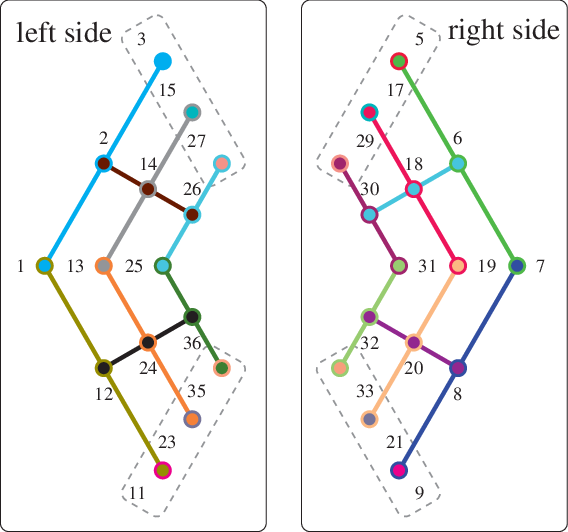}
}
\end{center}
\caption{\label{2023-navara-svozil-Rogalewicz-cut} The hypergraph depicted in Figure~\ref{2023-navara-svozil-Rogalewicz}
`cut' along the `vertical axis' formed by $\{4,16,28\}$ and $\{10,22,34\}$.
}
\end{figure}


As depicted in Figure~\ref{2023-navara-svozil-Rogalewicz-cut} the two parts are the gadgets from Figure~\ref{2023-navara-svozil-Rogalewicz-small} and we use their coordinatization from the previous section.
On the left-hand side, we utilize it in a literal sense, with the angle $\alpha$ serving as a degree of freedom.
The right-hand side is a mirror image of the left-hand side, and we employ the same coordinatization (with the same $\alpha$). The only difference is that we rotate it around the $z$-axis by an angle $\beta$.
This means that `adjacent' pairs of points on the left and the right side,
such as the pairs $2$ and $6$ or $8$ and $12$ or, in particular,  $3$ and $5$, are a rotation angle $\beta$ along the $z$-axis apart.
Moreover, because of this rotation around the $z$-axis, points $3$ and $5$ lie on the same circle with the same `longitudinal' $z$-coordinate.

In the final stage, the two parts will be `pasted' or `stitched' together by observing the proper orthogonality relations.
To achieve this, $\beta$ must be chosen in such a way that points $3$ and $5$ become orthogonal. Once this condition is met, all other necessary orthogonalities follow from the rotational symmetry.

To determine $\beta$ as a function of $\alpha$,
we first observe that,  as has been pointed out before, points $3$ and $5$ lie on the same circle with the same longitudinal $z$-coordinate.
For point $3$ this longitudinal $z$-coordinate is already determined by the construction mentioned earlier in Section~\ref{2023-navara-svozil-lsFORs}. More explicitly, in terms of the cross product of vectors in three dimensions, $\vert v_3 \rangle$ is the unit vector in the direction
$\vert v_2 \rangle \times \vert v_1 \rangle$.
Because
$\vert v_1 \rangle$
is the unit vector in the direction $\vert v_2 \rangle \times \vert v_{12} \rangle$,  we end up with $\vert v_3 \rangle$ being the unit vector in the direction $\vert v_2 \rangle \times ( \vert v_2 \rangle \times \vert v_{12} \rangle)$.

Therefore, given the longitudinal third coordinate of $\vert v_3 \rangle$ (which is dependent on $\alpha$) we already know the identical longitudinal third coordinate of the corresponding vector $\vert v_5 \rangle$.
All that is needed is to rotate $\vert v_5 \rangle$ by a rotation angle $\beta$ along the $z$-axis so that the scalar product with $\vert v_3 \rangle$ vanishes, that is, $\langle v_3\vert v_5 \rangle =0$. This yields $\beta$ as a function of $\alpha$:
\begin{equation}
\begin{split}
\beta (\alpha) =& \arccos\left(\frac{- 1 + \cos \alpha }{5+4\cos \alpha} \right)\\
=& \text{arcsec}\left[4 + 9\left(-1+\cos \alpha\right)^{-1}\right]
.
\end{split}
\label{2023-navara-svozil-betaofalpha}
\end{equation}
Note that the third longitudinal coordinates of $\vert v_3 \rangle$ and $\vert v_5 \rangle$ must be less than $1/\sqrt2$ because if they exceed $1/\sqrt2$, this circle cannot contain orthogonal vectors. In the case where the third longitudinal coordinate is $1/\sqrt2$, the orthogonal vectors are on opposite sides, resulting in $\beta=\pi$.
Some  values of $\beta$ as a function of $\alpha$ are presented in Table~\ref{t:alpha-beta}.

\begin{table}[ht]
\begin{tabular}{ll}
\toprule
    $\alpha                           $  & $\beta(\alpha) $  \\
\colrule
    $0                                $  & $\frac\pi 2 $   \\
    $\frac\pi 3                       $  & $\pi-\arccos\frac1{14} =\text{arcsec}(-14)$   \\
    $\frac23\pi                       $  & $\frac23\pi $   \\
    $\pi-\arccos \frac45  \quad$  & $\pi$ \\
\botrule
\end{tabular}
    \caption{Typical values of the function $\beta$ as a function of $\alpha$ in our construction. The value $\alpha=\beta = 2\pi/3$ is disallowed because it yields degeneracies due to multiplicities of vectors, see Section~\ref{2023-navara-svozil-lsFORs}.}
    \label{t:alpha-beta}
\end{table}

The function $\beta(\alpha)$ is defined for $\alpha\in[0,\alpha_{\mathrm{max}}]$,
where $\alpha_{\mathrm{max}}=\pi-\arccos (4/5)=2\arctan 3$.
In combination with the restrictions from the previous section, we allow values
$\alpha\in(0,2\arctan 3 ]\setminus\{2\pi /3, \alpha_0\}$.
In all these cases, we obtain (nonisomorphic) coordinatizations.
We acknowledge that there may be some (finitely many) values of $\alpha$ that could lead to additional orthogonalities not depicted in the hypergraph.

For example, for $\alpha =  \pi/3$ and $\beta\left( \pi/ 3\right) =\text{arcsec}(-14)$, the vector labels of the pseudocontexts are rendered by this construction as follows:


\begin{equation}
\begin{split}
\vert { \bf  v_{4}}  \rangle  & =    \left(  \frac{1}{14} \sqrt{\frac{1}{15} \left(209-9 \sqrt{65}\right)} , -\frac{5+3 \sqrt{65}}{70 \sqrt{2}} , -\sqrt{\frac{13}{15}} \right), \\
\vert {  \bf v_{16}} \rangle   &  =         \left(  \frac{\sqrt{65}-3}{14 \sqrt{6}} , \frac{1}{14} \sqrt{\frac{69}{5}+\sqrt{65}} , -\sqrt{\frac{13}{15}} \right), \\
\vert {  \bf v_{28}} \rangle  & =         \left(  -\frac{15+2 \sqrt{65}}{35 \sqrt{6}} , \frac{\sqrt{65}-10}{35 \sqrt{2}} , -\sqrt{\frac{13}{15}} \right), \\
\vert { \bf  v_{10}} \rangle   &  =   \left(  \frac{1}{\sqrt{30 \left(97+12 \sqrt{65}\right)}} , \frac{10+\sqrt{65}}{35 \sqrt{2}} , -\sqrt{\frac{13}{15}} \right), \\
\vert { \bf  v_{22}} \rangle  & =   \left(  -\frac{3+\sqrt{65}}{14 \sqrt{6}} , -\frac{1}{14} \sqrt{\frac{1}{5} \left(69-5 \sqrt{65}\right)} , -\sqrt{\frac{13}{15}} \right), \\
\vert { \bf   v_{34}} \rangle  &  =   \left(  \frac{45+\sqrt{65}}{70 \sqrt{6}} , \frac{5-3 \sqrt{65}}{70 \sqrt{2}} , -\sqrt{\frac{13}{15}} \right).
\end{split}
\end{equation}

Notice that all six lie in the plane $z=-\sqrt{ 13/15}$\,.
There is an equidistancing of the two triples
$\{4,16,28\}$
and
$\{10,22,34\}$ in terms of the Hilbert space inner products:
\begin{equation}
\begin{split}
&\langle v_{4} \vert v_{16}\rangle =
\langle v_{4} \vert v_{28}\rangle =
\langle v_{16} \vert v_{28}\rangle \\
&=\langle v_{10} \vert v_{22}\rangle =
\langle v_{10} \vert v_{34}\rangle =
\langle v_{22} \vert v_{34}\rangle =
\frac{4}{5} .
\end{split}
\end{equation}

The sums of the projection operators of the respective triples yield diagonal matrices
\begin{equation}
\begin{split}
&
\vert v_{4} \rangle \langle v_{4}  \vert +
\vert v_{16} \rangle \langle v_{16}\vert +
\vert v_{28} \rangle \langle v_{28}\vert
\\
&=
\vert v_{10}   \rangle \langle v_{10}\vert +
\vert v_{22}  \rangle \langle v_{22}\vert +
\vert v_{34}  \rangle \langle v_{34}\vert \\
&=
\text{diag} \left(\frac{1}{5}, \frac{1}{5}, \frac{13}{5}\right).
\end{split}
\end{equation}

A min-max argument~\cite{filipp-svo-04-qpoly-prl} yields bounds for the quantum probabilities
of the sums of observables in the two pseudocontexts $\{4, 16, 28\}$ and $\{10, 22, 34\}$ that are strictly smaller than the classical bounds (\ref{2023-navara-svozil-bftws}). In this case:
\begin{equation}
\begin{split}
0 < \frac{1}{5} \le & \; p(4) + p(16) + p(28) \\ &= p(10) + p(22) + p(34) \le \frac{13}{5} < 3.
\label{2023-navara-svozil-bfqp}
\end{split}
\end{equation}

The triples of vectors
$\{
 \vert   v_{4}  \rangle ,
 \vert   v_{16}  \rangle ,
 \vert   v_{28}  \rangle \}$
and $\{
 \vert   v_{10}  \rangle ,
 \vert   v_{22}  \rangle ,
 \vert   v_{34}  \rangle\}$
lie on a cone with the $z$-axis vector $(0,0,1)$ as its symmetry axis $\vert z \rangle = \left( 0,0,1 \right)$,
and an aperture of
$
\text{arccos}\vert \langle    v_{4}    \vert  z  \rangle \vert = \cdots = \text{arccos}\vert \langle    v_{z}    \vert  z  \rangle \vert
= \text{arccos} \sqrt{ 13/15} \approx
0.374  \equiv 21.4^\circ$.
That is, the construction renders vectors in the pseudocontexts whose convex combinations (with equal weights) are equal to the $z$-axis $(0,0,1)$.

\begin{figure}
\begin{center}
\includegraphics[width=.45\textwidth]{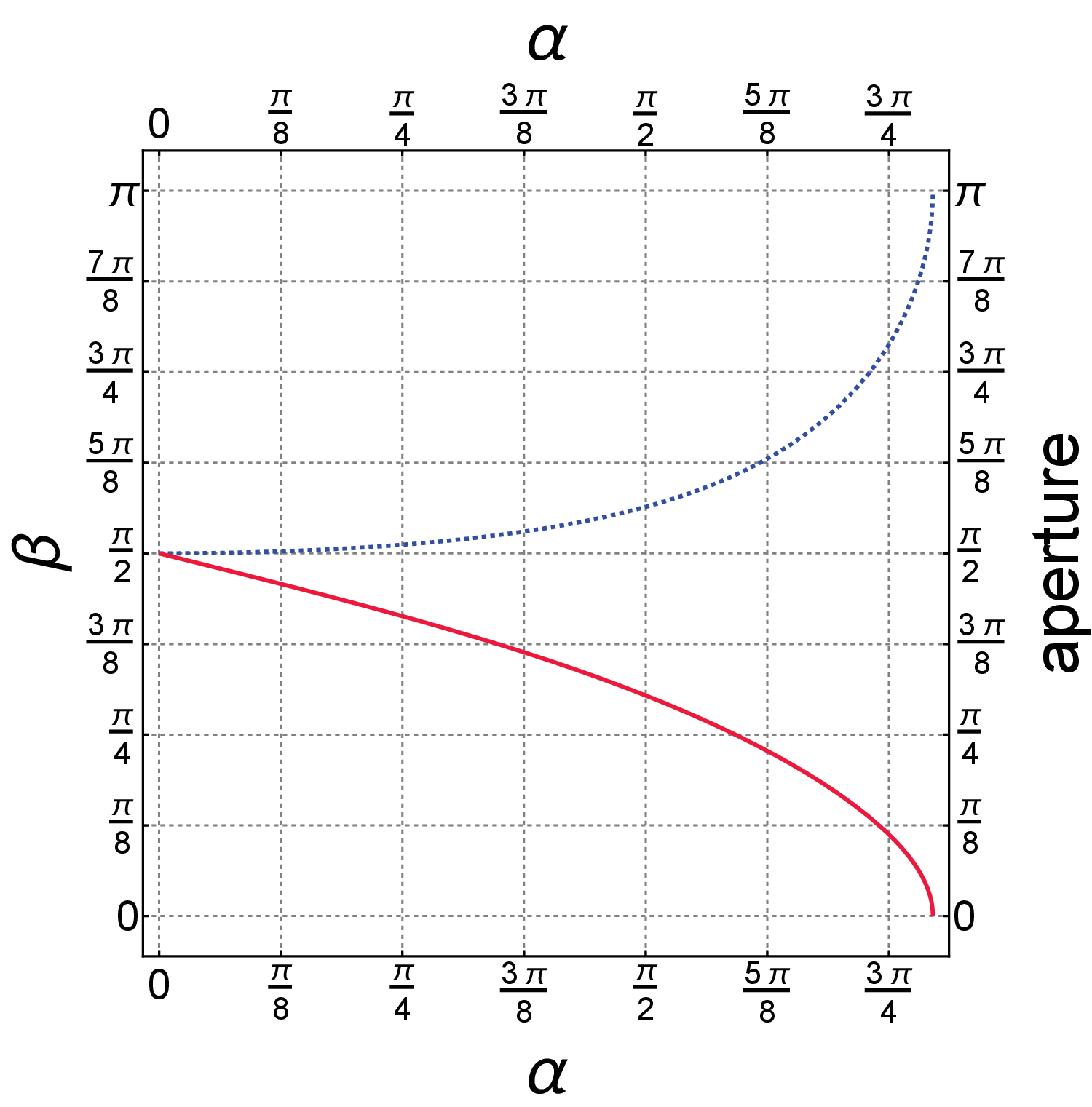}
\end{center}
\caption{\label{2023-navara-svozil-beta-aperture-versus-alpha}
Dependencies of the angle $\beta$ and the `aperture' angle of the cone, on which $\{4, 16, 28\}$ and $\{10, 22, 34\}$ lie relative to the $z$-axis. The dashed (blue) line represents $\beta$, and the solid (red) line represents the aperture angle as functions of the angle $\alpha$.
The units of angles are in radians.
}
\end{figure}

Figure~\ref{2023-navara-svozil-beta-aperture-versus-alpha} presents the dependencies of the angle $\beta$ between corresponding vectors on the contexts $\{2,14,26\}$ and $\{6,18,30\}$,
and the `aperture' angle
of the cone on which $\{4,16,28\}$ and $\{10,22,34\}$ lie relative to the $z$-axis,
as a function of the angle $\alpha$ between corresponding vectors on the contexts $\{2,14,26\}$ and $\{12,24,36\}$
as well as $\{6,18,30\}$ and $\{8,20,32\}$.

\section{Discussion}

The observed scenario can be understood within the framework of Spekkens' contextuality~\cite{Spekkens-04}, a broader concept that extends the initial understanding of contextuality within quantum theory. Measurement contextuality highlights differing statistical predictions across various models, particularly concerning quantized systems and classical ontological models.

The type of measurement contextuality introduced earlier is independent of the preparation procedure. We can establish equivalence between measurement procedures by summing the probabilities of elements within their corresponding pseudocontexts. This summation process enables us to define equivalence relations among measurements, resulting in the formation of distinct equivalence classes.

This form of contextuality arises from the observation that quantum realizations of the hypergraph defining the pseudocontexts,
which adhere to the equivalence constraints, do not violate a specific bound on the sum of probabilities within these pseudocontexts.
In contrast, classical realizations can exceed this bound.

This new type of contextuality can be quantified by observing the sums of the observables in the pseudocontexts such as $\{4, 16, 28\}$ and $\{10, 22, 34\}$
depicted in Figure~\ref{2023-navara-svozil-Rogalewicz}.
As mentioned earlier, there exists a separating set of two-valued states that are either $0$ or, alternatively, $1$  on all of the observables in the pseudocontexts. Consequently, there exist classical bounds~(\ref{2023-navara-svozil-bftws}) that exceed the quantum bounds~(\ref{2023-navara-svozil-bfqp}) obtained from a min-max calculation.

Pseudocontexts in general, and the statistical equivalence of pseudocontexts in particular, have the potential to serve as valuable tools for studying quantized systems beyond merely certifying their non-classical nature. These pseudocontexts can be correlated with non-orthogonal bases in Hilbert space, and, due to the preparation independence of their probability sums, represent generalizations of orthogonal frames.

Equality of the probability sums across pseudocontexts is only possible when considering more than two observables (e.g., triples) within each pseudocontext and in Hilbert spaces of dimensions greater than two. This restriction arises from the inherent impossibility of intertwining edges with fewer than three observables.

Assume two couples of vectors, $\{|a\rangle, |b\rangle\}$, $\{|c\rangle, |d\rangle\}$, such that each state attains the same sum of values on them.
In particular, this holds for any vector state determined by a vector $|e\rangle$.
If $|e\rangle$ is orthogonal to $|a\rangle, |b\rangle$, this sum is zero, hence $|e\rangle$ must be orthogonal also to $|c\rangle, |d\rangle$.
This implies the equality of the orthogonal spaces, $\{|a\rangle, |b\rangle\}^\perp=\{|c\rangle, |d\rangle\}^\perp$, and all four vectors $|a\rangle, |b\rangle, |c\rangle, |d\rangle$ lie in a plane.
Taking now $|e\rangle$ in this plane, the sums of the corresponding vector state over the couples are not constant because $|a\rangle \not\perp |b\rangle$.
The maximum is attained when $|e\rangle$ is an axis of symmetry of $|a\rangle, |b\rangle$
and the value of this maximum determines the angle of $|a\rangle, |b\rangle$.
The same applies to $|c\rangle, |d\rangle$, so these couples must be equal.

Furthermore, these pseudocontexts can also function as gadgets for both false-implies-false and true-implies-true scenarios.
They constitute broader and extended variations of the Specker bug and hypergraphs that necessitate a true-implies-true (TITS) set of two-valued states.

It can be argued that, to a certain extent, these observables are intricately interconnected or grouped together,
despite lacking mutual co-measurability. This situation draws parallels with the
Einstein-Podolski-Rosen scenario~\cite{epr,Howard1985171}.
For instance, one could consider the utilization of an entangled singlet
state involving two or three constituents (in three dimensions per constituent)~\cite[Table~4]{schimpf-svozil},
simultaneously measuring all elements of a pseudocontext, or alternatively, elements in different pseudocontexts,
$(1/\sqrt{3})\big(-\vert 0,0\rangle+\vert -1,1\rangle+\vert 1,-1\rangle\big)$ and
$-(1/\sqrt{6})\big(\vert -1,0,1\rangle+\vert 0,1,-1\rangle+\vert 1,-1,0\rangle\big)+
 (1/\sqrt{6})\big(\vert -1,1,0\rangle+\vert 0,-1,1\rangle+\vert 1,0,-1\rangle\big)$, respectively.
For instance, with a two-partite singlet state and in an Einstein-Podolsky-Rosen-type setup, we can measure one observable of one pseudocontext with one of the two particles. Similarly, we can measure one observable of the other pseudocontext with the second particle in the entangled particle pair.
Observing the cumulative statistics,
potentially employing a protocol akin to Bennett and Brassard's approach~\cite{benn-84},
could offer a dependable means of certifying the generation of quantum random numbers.

\begin{acknowledgments}
We are grateful to Josef Tkadlec for providing a {\em Pascal} program that computes and analyses the set of two-valued states of collections of contexts.
We are also grateful to  Norman D. Megill and Mladen Pavi{\v{c}}i{\'{c}} for providing a {\em C++} program that heuristically computes the faithful orthogonal representations of hypergraphs written in MMP format, given possible vector components.

This research was funded in whole, or in part, by the Austrian Science Fund (FWF), Project No. I 4579-N,
and by the Czech Science Foundation grant 20-09869L.

The authors declare no conflict of interest.
\end{acknowledgments}

\bibliography{svozil}
\bibliographystyle{apsrev}

\newpage

\begin{widetext}
\appendix

\section{Boolean set representation of the logic depicted in Figures~\ref{2023-navara-svozil-Rogalewicz-small}
and~\ref{2023-navara-svozil-Rogalewicz}}

`Bare, generic' labels $1,\ldots ,15$ and  $1,\ldots ,36$ are used for the exposition of the original hypergraph. Different label types $b_1,\ldots ,b_{36}$ are used for set representations by partition logics, and $\vert v_1\rangle ,\ldots ,\vert v_{36} \rangle$ for vector label representations.

A systematic way of generating a Boolean set representation is by computing all two-valued states of all atomic propositions, and then,
for each of the atoms,
generating an index set of all those two-valued states acquiring the value `1' on that atom~\cite{svozil-2001-eua}.

\subsection{Boolean set representation of the logic depicted in Figure~\ref{2023-navara-svozil-Rogalewicz-small}}

For the the logic depicted in Figure~\ref{2023-navara-svozil-Rogalewicz-small} the 24 two-valued states yield the following quasiclassical vertex labels:

\begin{equation}
\begin{split}
{ \bf     b_{1}}  =& \{1,2,3,4,5,6,7,8,9,10\},        \\
  b_{2}  =& \{11,12,13,14,15,16,17,18\},                          \\
  b_{3}  =& \{19,20,21,22,23,24\},                                \\
  b_{4}  =& \{1,2,3,4,11,12,13,14\},                              \\
{ \bf     b_{5}}  =& \{5,6,7,8,9,10,15,16,17,18\},   \\
{ \bf     b_{6}}  =& \{1,5,6,11,12,15,16,17,19,20\},  \\
  b_{7}  =& \{2,3,7,8,9,21,22,23\},                               \\
  b_{8}  =& \{4,10,13,14,18,24\},                                 \\
  b_{9}  =& \{5,7,8,15,16,19,21,22\},                             \\
{ \bf   b_{10}}  =& \{1,2,3,6,9,11,12,17,20,23\},    \\
{ \bf     b_{11}}  =& \{2,7,9,11,13,15,17,18,21,23\}, \\
  b_{12}  =& \{1,4,5,6,10,19,20,24\},                             \\
  b_{13}  =& \{3,8,12,14,16,22\},                                 \\
  b_{14}  =& \{6,9,10,17,18,20,23,24\},                           \\
{ \bf   b_{15}}  =& \{1,2,4,5,7,11,13,15,19,21\}
\end{split}
\end{equation}

\newpage

\subsection{Boolean set representation of the logic depicted in Figure~\ref{2023-navara-svozil-Rogalewicz}}

For the the logic depicted in Figure~\ref{2023-navara-svozil-Rogalewicz} the 225 two-valued states yield the following quasiclassical vertex labels:
\begin{equation}
\begin{split}
  b_{1}  =& \{1,2,3,4,5,6,7,8,9,10,11,12,13,14,15,16,17,18,19,20,21,22,23,24,25,26,27,28,29,30,       \\
& \qquad 31,32,33,34,35,36,37,38,39,40,41,42,43,44,45,46,47,48,49,50,51,52,53,54,55,56,57,58,59,60,61,  \\
& \qquad 62,63,64,65,66,67,68,69,70,71,72,73,74,75,76,77,78,79,80,81,82,83,84,85,86\}, \\
  b_{2}  =& \{87,88,89,90,91,92,93,94,95,96,97,98,99,100,101,102,103,104,105,106,107,108,109,         \\
& \qquad 110,111,112,113,114,115,116,117,118,119,120,121,122,123,124,125,126,127,128,129,130,131,132,   \\
& \qquad 133,134,135,136,137,138,139,140,141,142,143,144,145,146,147,148,149,150,151,152,153,154,       \\
& \qquad 155,156,157,158,159,160,161\}, \\
  b_{3}  =& \{162,163,164,165,166,167,168,169,170,171,172,173,174,175,176,177,178,179,                \\
& \qquad 180,181,182,183,184,185,186,187,188,189,190,191,192,193,194,195,196,197,198,199,200,201,       \\
& \qquad 202,203,204,205,206,207,208,209,210,211,212,213,214,215,216,217,218,219,220,                   \\
& \qquad 221,222,223,224,225\}, \\
{ \bf   b_{4}}  =& \{1,2,3,4,5,6,7,8,9,10,11,12,13,14,15,16,17,18,19,20,21,22,23,24,25,26,27,28,29,30,       \\
& \qquad 31,32,33,34,35,36,37,38,39,40,41,42,43,44,45,46,47,48,49,50,87,88,89,90,91,92,93,94,95,        \\
& \qquad 96,97,98,99,100,101,102,103,104,105,106,107,108,109,110,111,112,113,114,115,116,117,118,       \\
& \qquad 119,120,121,122,123,124,125,126,127,128,129,130,131,132,133\}, \\
  b_{5}  =& \{51,52,53,54,55,56,57,58,59,60,61,62,63,64,65,66,67,68,69,70,71,72,73,74,                \\
& \qquad 75,76,77,78,79,80,81,82,83,84,85,86,134,135,136,137,138,139,140,141,142,143,144,145,           \\
& \qquad 146,147,148,149,150,151,152,153,154,155,156,157,158,159,160,161\}, \\
  b_{6}  =& \{1,2,3,4,5,6,7,8,9,10,11,12,13,14,15,16,17,18,19,20,21,22,23,24,25,26,                   \\
& \qquad 87,88,89,90,91,92,93,94,95,96,97,98,99,100,101,102,103,104,105,106,107,162,163,164,            \\
& \qquad 165,166,167,168,169,170,171,172,173,174,175,176,177,178,179,180,181,182,183,184,               \\
& \qquad 185,186,187,188,189\}, \\
  b_{7}  =& \{27,28,29,30,31,32,33,34,35,36,37,38,39,40,41,42,43,44,45,46,47,48,                      \\
& \qquad 49,50,108,109,110,111,112,113,114,115,116,117,118,119,120,121,122,123,124,125,                 \\
& \qquad 126,127,128,129,130,131,132,133,190,191,192,193,194,195,196,197,198,199,200,201,               \\
& \qquad 202,203,204,205,206,207,208,209,210,211,212,213,214,215,216,217,218,219,220,221,               \\
& \qquad 222,223,224,225\}, \\
  b_{8}  =& \{1,2,3,4,5,6,7,8,9,10,11,12,13,14,51,52,53,54,55,56,57,58,59,60,61,                      \\
& \qquad 62,87,88,89,90,91,92,93,94,95,96,97,98,99,100,101,134,135,136,137,138,139,140,141,             \\
& \qquad 142,143,144,145,146,147,162,163,164,165,166,167,168,169,170,171,172,173,174,175,               \\
& \qquad 176,177,178,179,180,181\}, \\
  b_{9}  =& \{15,16,17,18,19,20,21,22,23,24,25,26,63,64,65,66,67,68,69,70,71,72,                      \\
& \qquad 73,74,75,76,77,78,79,80,81,82,83,84,85,86,102,103,104,105,106,107,148,149,150,                 \\
& \qquad 151,152,153,154,155,156,157,158,159,160,161,182,183,184,185,186,187,188,189\}, \\
{ \bf    b_{10} }=& \{1,2,3,4,5,6,7,8,9,10,11,12,13,14,27,28,29,30,31,32,33,34,35,36,                         \\
& \qquad 37,38,39,40,41,42,43,44,45,46,47,48,49,50,51,52,53,54,55,56,57,58,59,60,61,62,                 \\
& \qquad 87,88,89,90,91,92,93,94,95,108,109,110,111,112,113,114,115,116,117,118,119,                    \\
& \qquad 120,121,134,135,136,137,138,139,162,163,164,165,166,167,190,191,192,193,194,                   \\
& \qquad 195,196,197,198,199,200,201\}, \\
\end{split}
\end{equation}

\newpage

\begin{equation} \nonumber
\begin{split}
  b_{11} =& \{96,97,98,99,100,101,122,123,124,125,126,127,128,129,130,131,132,                        \\
& \qquad 133,140,141,142,143,144,145,146,147,168,169,170,171,172,173,174,175,176,177,                   \\
& \qquad 178,179,180,181,202,203,204,205,206,207,208,209,210,211,212,213,214,215,216,                   \\
& \qquad 217,218,219,220,221,222,223,224,225\},       \\
  b_{12} =& \{87,88,89,90,91,92,93,94,95,102,103,104,105,106,107,108,109,110,                         \\
& \qquad 111,112,113,114,115,116,117,118,119,120,121,134,135,136,137,138,139,148,149,                   \\
& \qquad 150,151,152,153,154,155,156,157,158,159,160,161,162,163,164,165,166,167,182,                   \\
& \qquad 183,184,185,186,187,188,189,190,191,192,193,194,195,196,197,198,199,200,201\}, \\
  b_{13} =& \{1,2,3,4,15,16,17,18,27,28,29,30,31,32,51,52,53,54,63,64,65,66,                          \\
& \qquad 67,68,87,88,89,90,91,92,96,97,102,103,104,105,106,108,109,110,111,112,113,114,                 \\
& \qquad 115,116,122,123,124,125,134,135,136,137,138,140,141,142,148,149,150,151,152,                   \\
& \qquad 153,154,155,156,162,163,168,169,170,171,182,183,184,190,191,192,193,202,203,                   \\
& \qquad 204,205,206,207\}, \\
  b_{14} =& \{5,6,7,8,9,10,11,19,20,21,22,23,24,33,34,35,36,37,38,39,40,41,42,                        \\
& \qquad 43,44,55,56,57,58,59,60,69,70,71,72,73,74,75,76,77,78,79,80,164,165,166,172,                   \\
& \qquad 173,174,175,176,177,178,185,186,187,188,194,195,196,197,198,199,208,209,210,                   \\
& \qquad 211,212,213,214,215,216,217,218,219\}, \\
  b_{15} =& \{12,13,14,25,26,45,46,47,48,49,50,61,62,81,82,83,84,85,86,93,94,                         \\
& \qquad 95,98,99,100,101,107,117,118,119,120,121,126,127,128,129,130,131,132,133,139,                  \\
& \qquad 143,144,145,146,147,157,158,159,160,161,167,179,180,181,189,200,201,220,221,                   \\
& \qquad 222,223,224,225\}, \\
{ \bf    b_{16}} =& \{1,2,5,6,7,8,15,19,27,28,29,30,33,34,35,36,37,38,39,40,51,52,53,                         \\
& \qquad 55,56,57,58,59,63,64,65,66,69,70,71,72,73,74,75,76,87,88,89,96,102,108,109,110,                \\
& \qquad 111,112,113,122,123,134,135,136,137,140,141,148,149,150,151,152,153,162,164,165,               \\
& \qquad 166,168,169,172,173,174,175,182,185,190,191,192,194,195,196,197,198,202,203,                   \\
& \qquad 204,205,208,209,210,211,212,213,214,215\}, \\
  b_{17} =& \{3,4,9,10,11,16,17,18,20,21,22,23,24,31,32,41,42,43,44,54,60,67,                         \\
& \qquad 68,77,78,79,80,90,91,92,97,103,104,105,106,114,115,116,124,125,138,142,154,155,                \\
& \qquad 156,163,170,171,176,177,178,183,184,186,187,188,193,199,206,207,216,217,218,219\}, \\
  b_{18} =& \{27,28,29,33,34,35,36,37,45,46,47,48,51,52,55,56,57,61,63,64,65,                         \\
& \qquad 69,70,71,72,73,81,82,83,84,108,109,110,111,117,118,119,122,126,127,128,129,130,                \\
& \qquad 134,135,136,140,143,144,145,148,149,150,151,157,158,159,190,191,194,195,196,200,               \\
& \qquad 202,203,204,208,209,210,211,212,220,221,222,223\}, \\
  b_{19} =& \{1,2,5,6,7,8,12,13,14,15,19,25,26,30,38,39,40,49,50,53,58,59,62,                         \\
& \qquad 66,74,75,76,85,86,87,88,89,93,94,95,96,98,99,100,101,102,107,112,113,120,121,                  \\
& \qquad 123,131,132,133,137,139,141,146,147,152,153,160,161,162,164,165,166,167,168,169,               \\
& \qquad 172,173,174,175,179,180,181,182,185,189,192,197,198,201,205,213,214,215,224,225\}, \\
  b_{20} =& \{16,17,20,21,22,23,27,28,31,33,34,35,36,41,42,45,46,47,63,64,67,69,                      \\
& \qquad 70,71,72,77,78,81,82,83,103,104,105,108,109,110,114,117,118,119,122,124,126,127,               \\
& \qquad 128,129,148,149,150,154,157,158,159,183,186,187,188,190,194,195,196,199,200,202,               \\
& \qquad 203,206,208,209,210,211,216,217,220,221,222\}, \\
\end{split}
\end{equation}

\newpage

\begin{equation} \nonumber
\begin{split}
  b_{21} =& \{3,4,9,10,11,18,24,29,32,37,43,44,48,51,52,54,55,56,57,60,61,65,68,                      \\
& \qquad 73,79,80,84,90,91,92,97,106,111,115,116,125,130,134,135,136,138,140,142,143,144,               \\
& \qquad 145,151,155,156,163,170,171,176,177,178,184,191,193,204,207,212,218,219,223\}, \\
{  \bf    b_{22}} =& \{1,2,5,6,7,12,15,16,17,19,20,21,22,25,27,28,30,31,33,34,35,38,39,41,                     \\
& \qquad 45,49,53,58,63,64,66,67,69,70,71,74,75,77,81,85,87,88,89,96,98,99,100,102,103,104,             \\
& \qquad 105,108,109,110,112,113,114,122,123,124,126,127,128,131,132,137,141,146,148,149,               \\
& \qquad 150,152,153,154,162,168,169,172,173,174,179,182,183,190,192,202,203,205,206,208,               \\
& \qquad 209,210,213,214,216,220,224\}, \\
  b_{23} =& \{8,13,14,23,26,36,40,42,46,47,50,59,62,72,76,78,82,83,86,93,94,95,                       \\
& \qquad 101,107,117,118,119,120,121,129,133,139,147,157,158,159,160,161,164,165,166,167,               \\
& \qquad 175,180,181,185,186,187,188,189,194,195,196,197,198,199,200,201,211,215,217,221,               \\
& \qquad 222,225\},                                                                                     \\
  b_{24} =& \{5,6,7,9,10,11,12,19,20,21,22,24,25,33,34,35,37,38,39,41,43,44,45,48,                    \\
& \qquad 49,55,56,57,58,60,61,69,70,71,73,74,75,77,79,80,81,84,85,98,99,100,126,127,128,130,            \\
& \qquad 131,132,143,144,145,146,172,173,174,176,177,178,179,208,209,210,212,213,214,216,218,           \\
& \qquad 219,220,223,224\}, \\
  b_{25} =& \{5,6,9,10,19,20,21,24,33,34,37,38,41,43,55,56,58,60,69,70,73,74,77,79,                   \\
& \qquad 87,88,90,91,93,94,98,99,102,103,104,106,107,108,109,111,112,114,115,117,118,120,126,           \\
& \qquad 127,130,131,134,135,137,138,139,143,144,146,148,149,151,152,154,155,157,158,160,164,           \\
& \qquad 165,172,173,176,177,185,186,187,194,195,197,199,208,209,212,213,216,218\}, \\
  b_{26} =& \{1,2,3,4,12,13,14,15,16,17,18,25,26,27,28,29,30,31,32,45,46,47,48,                       \\
& \qquad 49,50,51,52,53,54,61,62,63,64,65,66,67,68,81,82,83,84,85,86,162,163,167,168,169,               \\
& \qquad 170,171,179,180,181,182,183,184,189,190,191,192,193,200,201,202,203,204,205,206,               \\
& \qquad 207,220,221,222,223,224,225\}, \\
  b_{27} =& \{7,8,11,22,23,35,36,39,40,42,44,57,59,71,72,75,76,78,80,89,92,95,96,                     \\
& \qquad 97,100,101,105,110,113,116,119,121,122,123,124,125,128,129,132,133,136,140,141,                \\
& \qquad 142,145,147,150,153,156,159,161,166,174,175,178,188,196,198,210,211,214,215,217,219\}, \\
{ \bf    b_{28}} =& \{1,3,5,9,12,13,16,20,27,30,31,32,33,38,41,43,45,46,49,50,51,53,54,55,58,                 \\
& \qquad 60,61,62,63,66,67,68,69,74,77,79,81,82,85,86,87,90,93,98,103,108,112,114,115,117,120,          \\
& \qquad 126,131,134,137,138,139,143,146,148,152,154,155,157,160,162,163,164,167,168,170,172,           \\
& \qquad 176,179,180,183,186,190,192,193,194,197,199,200,201,202,205,206,207,208,213,216,218,           \\
& \qquad 220,221,224,225\}, \\
  b_{29} =& \{2,4,6,10,14,15,17,18,19,21,24,25,26,28,29,34,37,47,48,52,56,64,65,70,                   \\
& \qquad 73,83,84,88,91,94,99,102,104,106,107,109,111,118,127,130,135,144,149,151,158,165,169,          \\
& \qquad 171,173,177,181,182,184,185,187,189,191,195,203,204,209,212,222,223\}, \\
  b_{30} =& \{30,31,32,38,39,40,41,42,43,44,49,50,53,54,58,59,60,62,66,67,68,74,75,76,                \\
& \qquad 77,78,79,80,85,86,112,113,114,115,116,120,121,123,124,125,131,132,133,137,138,139,141,         \\
& \qquad 142,146,147,152,153,154,155,156,160,161,192,193,197,198,199,201,205,206,207,213,214,           \\
& \qquad 215,216,217,218,219,224,225\}, \\
\end{split}
\end{equation}

\newpage

\begin{equation}  \nonumber
\begin{split}
  b_{31} =& \{1,3,5,7,8,9,11,12,13,16,20,22,23,27,33,35,36,45,46,51,55,57,61,63,69,71,                \\
& \qquad 72,81,82,87,89,90,92,93,95,96,97,98,100,101,103,105,108,110,117,119,122,126,128,129,           \\
& \qquad 134,136,140,143,145,148,150,157,159,162,163,164,166,167,168,170,172,174,175,176,178,           \\
& \qquad 179,180,183,186,188,190,194,196,200,202,208,210,211,220,221\}, \\
  b_{32} =& \{15,18,19,24,25,26,29,30,32,37,38,39,40,43,44,48,49,50,65,66,68,73,74,                   \\
& \qquad 75,76,79,80,84,85,86,102,106,107,111,112,113,115,116,120,121,123,125,130,131,132,133,          \\
& \qquad 151,152,153,155,156,160,161,182,184,185,189,191,192,193,197,198,201,204,205,207,212,           \\
& \qquad 213,214,215,218,219,223,224,225\}, \\
  b_{33} =& \{2,4,6,10,14,17,21,28,31,34,41,42,47,52,53,54,56,58,59,60,62,64,67,70,77,                \\
& \qquad 78,83,88,91,94,99,104,109,114,118,124,127,135,137,138,139,141,142,144,146,147,149,154,         \\
& \qquad 158,165,169,171,173,177,181,187,195,199,203,206,209,216,217,222\}, \\
{ \bf    b_{34}} =& \{1,3,5,8,9,13,15,16,18,19,20,23,24,26,27,29,30,32,33,36,37,38,40,43,46,                  \\
& \qquad 50,51,55,63,65,66,68,69,72,73,74,76,79,82,86,87,90,93,96,97,98,101,102,103,106,107,            \\
& \qquad 108,111,112,115,117,120,122,123,125,126,129,130,131,133,134,140,143,148,151,152,155,           \\
& \qquad 157,160,164,168,170,172,175,176,180,185,186,194,197,202,204,205,207,208,211,212,213,           \\
& \qquad 215,218,221,225\}, \\
  b_{35} =& \{7,11,12,22,25,35,39,44,45,48,49,57,61,71,75,80,81,84,85,89,92,95,100,                   \\
& \qquad 105,110,113,116,119,121,128,132,136,145,150,153,156,159,161,162,163,166,167,174,178,           \\
& \qquad 179,182,183,184,188,189,190,191,192,193,196,198,200,201,210,214,219,220,223,224\}, \\
  b_{36} =& \{1,2,3,4,8,13,14,15,16,17,18,23,26,27,28,29,30,31,32,36,40,42,46,47,50,                  \\
& \qquad 51,52,53,54,59,62,63,64,65,66,67,68,72,76,78,82,83,86,96,97,101,122,123,124,125,129,           \\
& \qquad 133,140,141,142,147,168,169,170,171,175,180,181,202,203,204,205,206,207,211,215,217,           \\
& \qquad 221,222,225\}.
\end{split}
\end{equation}

\section{Faithful orthogonal representation of the logic depicted in Figure~\ref{2023-navara-svozil-Rogalewicz}}

According to an `inverted' definition~\cite{Portillo-2015} inspired by Lov\'asz~\cite{lovasz-79},
a faithful orthogonal representation or coordinatization of a hypergraph $G$ with elements $1, \ldots , n$
is a corresponding system of vector labels---that is, unit vectors ${ \vert v_1 \rangle , \ldots , \vert v_n \rangle }$ in a Euclidean space---such that
if $i$ and $j$ are in the same hyperedge,
then $\vert v_i \rangle$ and $\vert v_j \rangle$ are orthogonal.

There is currently no tractable systematic method for the coordinatization of hypergraphs.
The following two faithful orthogonal representations have been obtained by an ad hoc analytical approach,
using rotations by two angles $\alpha$ and $\beta$ along a common axis.

\subsection{$\alpha=\frac{\pi}{3}$ and $\beta\left(\frac{\pi}{3}\right) =\text{arcsec}(-14)$}

In the first faithful orthogonal representation of the hypergraph depicted in Figure~\ref{2023-navara-svozil-Rogalewicz},
 the values $\alpha=\frac{\pi}{3}$ and $\beta(\frac{\pi}{3}) =\text{arcsec}(-14)$ have been chosen.

\begin{equation}
\begin{split}
\vert v_{1}  \rangle  & =                        \left(  \sqrt{\frac{3}{10}} , -\frac{1}{\sqrt{10}} , -\sqrt{\frac{3}{5}} \right), \\
\vert v_{2}  \rangle  & =                        \left(  \sqrt{\frac{2}{3}} , 0 , \frac{1}{\sqrt{3}} \right), \\
\vert v_{3}  \rangle  & =                        \left(  -\frac{1}{\sqrt{30}} , -\frac{3}{\sqrt{10}} , \frac{1}{\sqrt{15}} \right), \\
\vert { \bf   v_{4}}  \rangle  & =    \left(  \frac{1}{14} \sqrt{\frac{1}{15} \left(209-9 \sqrt{65}\right)} , -\frac{5+3 \sqrt{65}}{70 \sqrt{2}} , -\sqrt{\frac{13}{15}} \right), \\
\vert v_{5}  \rangle  & =                        \left(  \frac{1}{420} \left(-\sqrt{30}-45 \sqrt{78}\right) , \frac{1}{14} \sqrt{\frac{1}{5} \left(37-3 \sqrt{65}\right)} , -\frac{1}{\sqrt{15}} \right), \\
\vert v_{6}  \rangle  & =                        \left(  -\frac{1}{7 \sqrt{6}} , \frac{\sqrt{\frac{65}{2}}}{7} , \frac{1}{\sqrt{3}} \right), \\
\vert v_{7}  \rangle  & =                        \left(  \frac{1}{14} \sqrt{\frac{3}{10}} \left(\sqrt{65}-1\right) , \frac{1+3 \sqrt{65}}{14 \sqrt{10}} , -\sqrt{\frac{3}{5}} \right), \\
\vert v_{8}  \rangle  & =                        \left(  \frac{3 \sqrt{65}-1}{14 \sqrt{6}} , \frac{\sqrt{33+\sqrt{65}}}{14} , \frac{1}{\sqrt{3}} \right), \\
\vert v_{9}  \rangle  & =                        \left(  \frac{1}{210} \left(2 \sqrt{30}+15 \sqrt{78}\right) , \frac{1}{70} \left(\sqrt{10}-10 \sqrt{26}\right) , -\frac{1}{\sqrt{15}} \right), \\
\vert { \bf  v_{10}} \rangle  & =   \left(  \frac{1}{\sqrt{30 \left(97+12 \sqrt{65}\right)}} , \frac{10+\sqrt{65}}{35 \sqrt{2}} , -\sqrt{\frac{13}{15}} \right), \\
\vert v_{11} \rangle  & =                        \left(  -2 \sqrt{\frac{2}{15}} , -\sqrt{\frac{2}{5}} , -\frac{1}{\sqrt{15}} \right), \\
\vert v_{12} \rangle  & =                        \left(  \frac{1}{\sqrt{6}} , -\frac{1}{\sqrt{2}} , \frac{1}{\sqrt{3}} \right), \\
\vert v_{13} \rangle  & =                        \left(  0 , \sqrt{\frac{2}{5}} , -\sqrt{\frac{3}{5}} \right), \\
\vert v_{14} \rangle  & =                        \left(  -\frac{1}{\sqrt{6}} , \frac{1}{\sqrt{2}} , \frac{1}{\sqrt{3}} \right), \\
\vert v_{15} \rangle  & =                        \left(  \sqrt{\frac{5}{6}} , \frac{1}{\sqrt{10}} , \frac{1}{\sqrt{15}} \right), \\
\vert {  \bf v_{16}} \rangle  & =         \left(  \frac{\sqrt{65}-3}{14 \sqrt{6}} , \frac{1}{14} \sqrt{\frac{69}{5}+\sqrt{65}} , -\sqrt{\frac{13}{15}} \right), \\
\vert v_{17} \rangle  & =                        \left(  \frac{1}{84} \left(\sqrt{30}+3 \sqrt{78}\right) , \frac{1}{140} \left(\sqrt{10}-25 \sqrt{26}\right) , -\frac{1}{\sqrt{15}} \right), \\
\vert v_{18} \rangle  & =                        \left(  \frac{1}{84} \left(\sqrt{6}-3 \sqrt{390}\right) , -\frac{1}{14} \sqrt{33+\sqrt{65}} , \frac{1}{\sqrt{3}} \right), \\
\end{split}
\end{equation}
\newpage

\begin{equation}  \nonumber
\begin{split}
\vert v_{19} \rangle  & =                        \left(  -\frac{\sqrt{\frac{39}{2}}}{7} , -\frac{1}{7 \sqrt{10}} , -\sqrt{\frac{3}{5}} \right), \\
\vert v_{20} \rangle  & =                        \left(  -\frac{1}{14} \sqrt{\frac{293}{3}+\sqrt{65}} , \frac{\sqrt{65}-1}{14 \sqrt{2}} , \frac{1}{\sqrt{3}} \right), \\
\vert v_{21} \rangle  & =                        \left(  \frac{1}{14} \sqrt{\frac{1}{3} \left(61-3 \sqrt{65}\right)} , \frac{1}{14} \sqrt{\frac{813}{5}+\sqrt{65}} , -\frac{1}{\sqrt{15}} \right), \\
\vert { \bf  v_{22}} \rangle  & =   \left(  -\frac{3+\sqrt{65}}{14 \sqrt{6}} , -\frac{1}{14} \sqrt{\frac{1}{5} \left(69-5 \sqrt{65}\right)} , -\sqrt{\frac{13}{15}} \right), \\
\vert v_{23} \rangle  & =                        \left(  \sqrt{\frac{5}{6}} , -\frac{1}{\sqrt{10}} , -\frac{1}{\sqrt{15}} \right), \\
\vert v_{24} \rangle  & =                        \left(  \frac{1}{\sqrt{6}} , \frac{1}{\sqrt{2}} , \frac{1}{\sqrt{3}} \right), \\
\vert v_{25} \rangle  & =                        \left(  -\sqrt{\frac{3}{10}} , -\frac{1}{\sqrt{10}} , -\sqrt{\frac{3}{5}} \right), \\
\vert v_{26} \rangle  & =                        \left(  -\frac{1}{\sqrt{6}} , -\frac{1}{\sqrt{2}} , \frac{1}{\sqrt{3}} \right), \\
\vert v_{27} \rangle  & =                        \left(  -2 \sqrt{\frac{2}{15}} , \sqrt{\frac{2}{5}} , \frac{1}{\sqrt{15}} \right), \\
\vert {  \bf  v_{28}} \rangle  & =         \left(  -\frac{15+2 \sqrt{65}}{35 \sqrt{6}} , \frac{\sqrt{65}-10}{35 \sqrt{2}} , -\sqrt{\frac{13}{15}} \right), \\
\vert v_{29} \rangle  & =                        \left(  \frac{1}{210} \left(15 \sqrt{78}-2 \sqrt{30}\right) , \frac{1}{70} \left(\sqrt{10}+10 \sqrt{26}\right) , -\frac{1}{\sqrt{15}} \right), \\
\vert v_{30} \rangle  & =                        \left(  \frac{1}{84} \left(\sqrt{6}+3 \sqrt{390}\right) , -\frac{\sqrt{65}-1}{14 \sqrt{2}} , \frac{1}{\sqrt{3}} \right), \\
\vert v_{31} \rangle  & =                        \left(  \frac{1}{14} \sqrt{\frac{3}{10}} \left(1+\sqrt{65}\right) , \frac{1-3 \sqrt{65}}{14 \sqrt{10}} , -\sqrt{\frac{3}{5}} \right), \\
\vert v_{32} \rangle  & =                        \left(  \frac{1}{7 \sqrt{6}} , -\frac{\sqrt{\frac{65}{2}}}{7} , \frac{1}{\sqrt{3}} \right), \\
\vert v_{33} \rangle  & =                        \left(  \frac{1}{420} \left(\sqrt{30}-45 \sqrt{78}\right) , -\frac{1}{14} \sqrt{\frac{1}{5} \left(37+3 \sqrt{65}\right)} , -\frac{1}{\sqrt{15}} \right), \\
\vert { \bf   v_{34}} \rangle & =   \left(  \frac{45+\sqrt{65}}{70 \sqrt{6}} , \frac{5-3 \sqrt{65}}{70 \sqrt{2}} , -\sqrt{\frac{13}{15}} \right), \\
\vert v_{35} \rangle  & =                        \left(  -\frac{1}{\sqrt{30}} , \frac{3}{\sqrt{10}} , -\frac{1}{\sqrt{15}} \right), \\
\vert v_{36} \rangle  & =                        \left(  -\sqrt{\frac{2}{3}} , 0 , \frac{1}{\sqrt{3}} \right).
\end{split}
\end{equation}

\newpage

\subsection{$\alpha=\frac{\pi}{2}$ and $\beta \left(\frac{\pi}{2}\right)= \text{arcsec}(-5)$}

The second faithful orthogonal representation of the hypergraph depicted in Figure~\ref{2023-navara-svozil-Rogalewicz}
uses the values $\alpha=\frac{\pi}{2}$ and $\beta \left(\frac{\pi}{2}\right)= \text{arcsec}(-5)$.

\begin{equation}
\begin{split}
\vert v_{1}  \rangle  & =                        \left( \frac{1}{2} , -\frac{1}{2} , -\frac{1}{\sqrt{2}} \right), \\
\vert v_{2}  \rangle  & =                        \left( \sqrt{\frac{2}{3}} , 0 , \frac{1}{\sqrt{3}} \right), \\
\vert v_{3}  \rangle  & =                        \left( -\frac{1}{2 \sqrt{3}} , -\frac{\sqrt{3}}{2} , \frac{1}{\sqrt{6}} \right), \\
\vert { \bf   v_{4}}  \rangle  & =   \left( \frac{1}{5} \sqrt{\frac{29}{6}+\sqrt{6}} , \frac{\sqrt{6}-1}{5 \sqrt{2}} , \sqrt{\frac{2}{3}} \right), \\
\vert v_{5}  \rangle  & =                        \left( \frac{1}{30} \left(18 \sqrt{2}-\sqrt{3}\right) , -\frac{1}{10} \sqrt{11+4 \sqrt{6}} , -\frac{1}{\sqrt{6}} \right), \\
\vert v_{6}  \rangle  & =                        \left( -\frac{\sqrt{\frac{2}{3}}}{5} , -\frac{4}{5} , \frac{1}{\sqrt{3}} \right), \\
\vert v_{7}  \rangle  & =                        \left( \frac{1}{10} \left(-1-2 \sqrt{6}\right) , \frac{1}{10} \left(1-2 \sqrt{6}\right) , -\frac{1}{\sqrt{2}} \right), \\
\vert v_{8}  \rangle  & =                        \left( -\frac{4}{5} , \frac{\sqrt{\frac{2}{3}}}{5} , \frac{1}{\sqrt{3}} \right), \\
\vert v_{9}  \rangle  & =                        \left( \frac{1}{10} \left(\sqrt{3}-2 \sqrt{2}\right) , \frac{1}{30} \left(18 \sqrt{2}+\sqrt{3}\right) , -\frac{1}{\sqrt{6}} \right), \\
\vert { \bf  v_{10}} \rangle  & =   \left( -\frac{1}{5} \sqrt{\frac{7}{2}+\sqrt{6}} , \frac{1}{5} \sqrt{\frac{29}{6}-\sqrt{6}} , \sqrt{\frac{2}{3}} \right), \\
\vert v_{11} \rangle  & =                        \left( -\frac{\sqrt{3}}{2} , -\frac{1}{2 \sqrt{3}} , -\frac{1}{\sqrt{6}} \right), \\
\vert v_{12} \rangle  & =                        \left( 0 , -\sqrt{\frac{2}{3}} , \frac{1}{\sqrt{3}} \right), \\
\vert v_{13} \rangle  & =                        \left( \frac{1}{4} \left(\sqrt{3}-1\right) , \frac{1}{4} \left(1+\sqrt{3}\right) , -\frac{1}{\sqrt{2}} \right), \\
\vert v_{14} \rangle  & =                        \left( -\frac{1}{\sqrt{6}} , \frac{1}{\sqrt{2}} , \frac{1}{\sqrt{3}} \right), \\
\vert v_{15} \rangle  & =                        \left( \frac{1}{12} \left(9+\sqrt{3}\right) , \frac{1}{4} \left(\sqrt{3}-1\right) , \frac{1}{\sqrt{6}} \right), \\
\vert {  \bf v_{16}} \rangle  & =    \left( \frac{1}{60} \left(-18-9 \sqrt{2}-2 \sqrt{3}+3 \sqrt{6}\right) , \frac{1}{20} \left(2+\sqrt{2}-2 \sqrt{3}+3 \sqrt{6}\right) , \sqrt{\frac{2}{3}} \right), \\
\vert v_{17} \rangle  & =                        \left( \frac{1}{60} \left(9-18 \sqrt{2}+\sqrt{3}+6 \sqrt{6}\right) , \frac{1}{20} \left(-1+2 \sqrt{2}+\sqrt{3}+6 \sqrt{6}\right) , -\frac{1}{\sqrt{6}} \right), \\
\vert v_{18} \rangle  & =                        \left( \frac{12+\sqrt{2}}{10 \sqrt{3}} , \frac{1}{10} \left(4-\sqrt{2}\right) , \frac{1}{\sqrt{3}} \right),
\end{split}
\end{equation}
\newpage

\begin{equation}  \nonumber
\begin{split}
\vert v_{19} \rangle  & =                        \left(  \frac{1}{20} \left(1+6 \sqrt{2}-\sqrt{3}+2 \sqrt{6}\right) , \frac{1}{20} \left(-1-6 \sqrt{2}-\sqrt{3}+2 \sqrt{6}\right) , -\frac{1}{\sqrt{2}} \right), \\
\vert v_{20} \rangle  & =                        \left(  \frac{1}{10} \left(4-\sqrt{2}\right) , -\frac{12+\sqrt{2}}{10 \sqrt{3}} , \frac{1}{\sqrt{3}} \right), \\
\vert v_{21} \rangle  & =                        \left(  \frac{1}{20} \left(-1-6 \sqrt{6}+\sqrt{11-4 \sqrt{6}}\right) , \frac{1}{60} \left(9-18 \sqrt{2}-\sqrt{3}-6 \sqrt{6}\right) , -\frac{1}{\sqrt{6}} \right), \\
\vert { \bf  v_{22}} \rangle  & =   \left(  \frac{1}{20} \left(2+\sqrt{2}+2 \sqrt{3}-3 \sqrt{6}\right) , \frac{1}{60} \left(-18-9 \sqrt{2}+2 \sqrt{3}-3 \sqrt{6}\right) , \sqrt{\frac{2}{3}} \right), \\
\vert v_{23} \rangle  & =                        \left(  \frac{1}{4} \left(1+\sqrt{3}\right) , \frac{1}{12} \left(\sqrt{3}-9\right) , -\frac{1}{\sqrt{6}} \right), \\
\vert v_{24} \rangle  & =                        \left(  \frac{1}{\sqrt{2}} , \frac{1}{\sqrt{6}} , \frac{1}{\sqrt{3}} \right), \\
\vert v_{25} \rangle  & =                        \left(  \frac{1}{4} \left(-1-\sqrt{3}\right) , \frac{1}{4} \left(1-\sqrt{3}\right) , -\frac{1}{\sqrt{2}} \right), \\
\vert v_{26} \rangle  & =                        \left(  -\frac{1}{\sqrt{6}} , -\frac{1}{\sqrt{2}} , \frac{1}{\sqrt{3}} \right), \\
\vert v_{27} \rangle  & =                        \left(  \frac{1}{12} \left(\sqrt{3}-9\right) , \frac{1}{4} \left(1+\sqrt{3}\right) , \frac{1}{\sqrt{6}} \right), \\
\vert {  \bf  v_{28}} \rangle  & =   \left(  \frac{1}{60} \left(18-9 \sqrt{2}-2 \sqrt{3}-3 \sqrt{6}\right) , \frac{1}{20} \left(-2+\sqrt{2}-2 \sqrt{3}-3 \sqrt{6}\right) , \sqrt{\frac{2}{3}} \right), \\
\vert v_{29} \rangle  & =                        \left(  \frac{1}{60} \left(-9-18 \sqrt{2}+\sqrt{3}-6 \sqrt{6}\right) , \frac{1}{20} \left(1+2 \sqrt{2}+\sqrt{3}-6 \sqrt{6}\right) , -\frac{1}{\sqrt{6}} \right), \\
\vert v_{30} \rangle  & =                        \left(  \frac{\sqrt{2}-12}{10 \sqrt{3}} , \frac{1}{10} \left(4+\sqrt{2}\right) , \frac{1}{\sqrt{3}} \right), \\
\vert v_{31} \rangle  & =                        \left(  \frac{1}{20} \left(1-6 \sqrt{2}+\sqrt{3}+2 \sqrt{6}\right) , \frac{1}{20} \left(-1+6 \sqrt{2}+\sqrt{3}+2 \sqrt{6}\right) , -\frac{1}{\sqrt{2}} \right), \\
\vert v_{32} \rangle  & =                        \left(  \frac{1}{10} \left(4+\sqrt{2}\right) , \frac{1}{5} \sqrt{\frac{73}{6}-2 \sqrt{2}} , \frac{1}{\sqrt{3}} \right), \\
\vert v_{33} \rangle  & =                        \left(  \frac{1}{20} \left(1+6 \sqrt{6}+\sqrt{11-4 \sqrt{6}}\right) , \frac{1}{60} \left(-9-18 \sqrt{2}-\sqrt{3}+6 \sqrt{6}\right) , -\frac{1}{\sqrt{6}} \right), \\
\vert { \bf   v_{34}} \rangle & =   \left(  \frac{1}{20} \left(-2+\sqrt{2}+2 \sqrt{3}+3 \sqrt{6}\right) , \frac{1}{60} \left(18-9 \sqrt{2}+2 \sqrt{3}+3 \sqrt{6}\right) , \sqrt{\frac{2}{3}} \right), \\
\vert v_{35} \rangle  & =                        \left(  \frac{1}{4} \left(\sqrt{3}-1\right) , \frac{1}{12} \left(9+\sqrt{3}\right) , -\frac{1}{\sqrt{6}} \right), \\
\vert v_{36} \rangle  & =                        \left(  -\frac{1}{\sqrt{2}} , \frac{1}{\sqrt{6}} , \frac{1}{\sqrt{3}} \right).
\end{split}
\end{equation}

\newpage
\end{widetext}

\end{document}